\documentclass{article}
\usepackage{spconf,amsmath,graphicx}
\usepackage{stfloats}
\usepackage{booktabs}
\usepackage{multirow}
\usepackage{bbding}
\usepackage{xcolor}
\usepackage{hyperref}
\usepackage{enumitem}
\usepackage[normalem]{ulem}

\title{Improving Contextual Spelling Correction by External Acoustics Attention and Semantic Aware Data Augmentation}
\name{Xiaoqiang Wang, Yanqing Liu, Jinyu Li, Sheng Zhao}
\address{Microsoft Corporation, Redmond, WA, US}
\begin{document}
\ninept
\maketitle
\begin{abstract}
We previously proposed contextual spelling correction (CSC) to correct the output of end-to-end (E2E) automatic speech recognition (ASR) models with contextual information such as name, place, etc. Although CSC has achieved reasonable improvement in the biasing problem, there are still two drawbacks for further accuracy improvement. 
First, due to information limitation in text only hypothesis or weak performance of ASR model on rare domains, the CSC model may fail to correct phrases with similar pronunciation or anti-context cases where all biasing phrases are not present in the utterance. Second, there is a discrepancy between the training and inference of CSC. The bias list in training is randomly selected but in inference there may be more similarity between ground truth phrase and other phrases.  
To solve above limitations, in this paper we propose an improved non-autoregressive (NAR) spelling correction model for contextual biasing in E2E neural transducer-based ASR systems to improve the previous CSC model from two perspectives: Firstly, we incorporate acoustics information with an external attention as well as text hypotheses into CSC to better distinguish target phrase from dissimilar or irrelevant phrases. Secondly, we design a semantic aware data augmentation schema in training phrase to reduce the mismatch between training and inference to further boost the biasing accuracy. Experiments show that the improved method outperforms the baseline ASR+Biasing system by as much as 20.3\% relative name recall gain and achieves stable improvement compared to the previous CSC method over different bias list name coverage ratio.

\end{abstract}
\section{Introduction}
\label{sec:intro}

Contextual biasing is a challenging task for end-to-end (E2E) automatic speech recognition (ASR) systems \cite{li2022recent}, which improves the recognition performance by biasing the model to a specific domain phrase list including a user's contact names, songs, location, and other contextual information. 
Prior works for contextual biasing of E2E ASR system can be classified into three categories. 
The first method is to represent the biasing phrases as a finite state transducer (FST) and incorporate it into the beam-search decoding framework of E2E model \cite{OTF_rescore2, OTF_rescore3, CLAS, OTF_rescore4, le2021deep}. 
The second method is to directly incorporate the contextual information into the E2E model with a bias encoder \cite{CLAS, jain2020contextual, phoebe, trie_deep_biasing, instant_one_shot, tree_constrained}. 
To deal with the scalability issues \cite{CLAS} when with large biasing phrase list and further improve the biasing performance, the contextual spelling correction method is also proposed by biasing the recognition hypothesis with an efficient and small contextual spelling correction (CSC) model \cite{CSC, wang2022towards}, which acts as a post processing module. 
Compared to the first two categories, due to the post processing nature, CSC can pre-select biasing phrases with a filter mechanism, which reduces the effective number of biasing phrases to avoid attention diffusion. 

\begin{figure*}[h]
  \centering
  \includegraphics[width=1.0\linewidth]{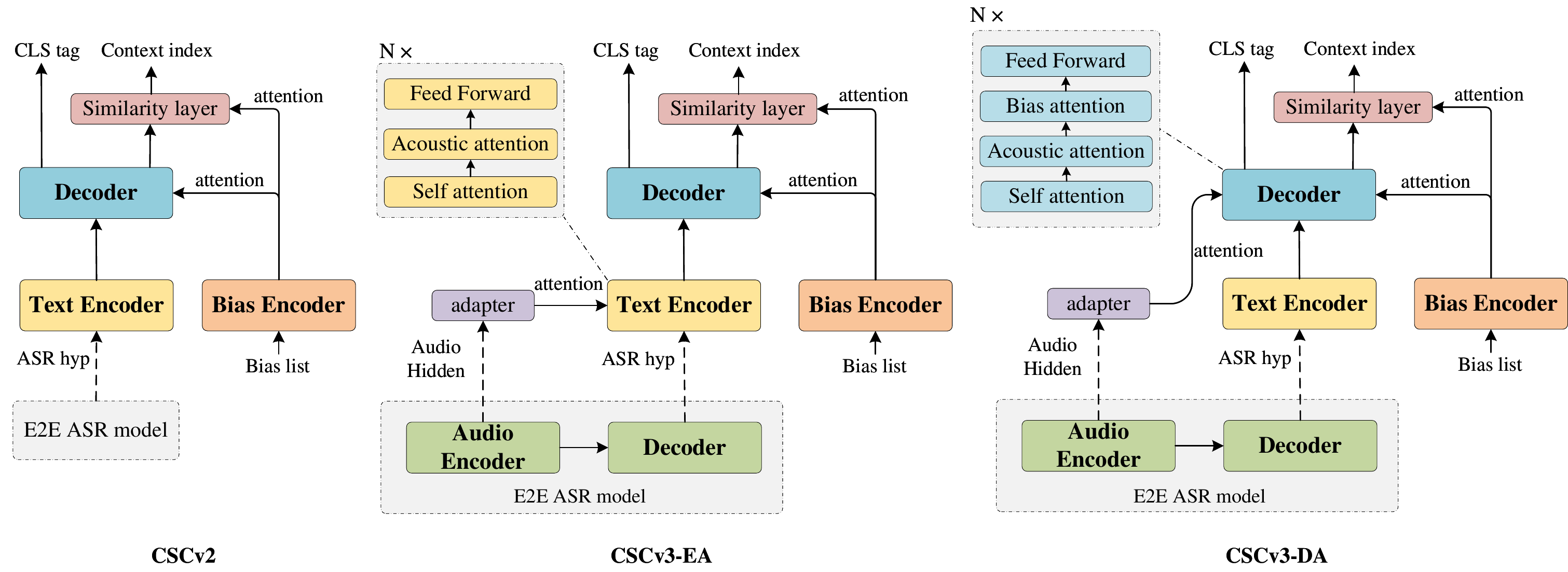}
  \caption{Model structure of CSCv2 and CSCv3. CSCv3 model leverages acoustic information with external acoustic attention. We propose two architectures: CSCv3-EA incorporates acoustic information in text encoder, while CSCv3-DA incorporates acoustic information in decoder.}
  \label{fig:csc_v3_model}
\end{figure*}

Autoregressive contextual spelling correction (CSCv1) \cite{CSC} on top of ASR model as a post processing method has shown improvement on phrase biasing problems, which incorporates the contextual information into an autoregressive (AR) spelling correction model \cite{SC_1,SC_4}, but the efficiency is poor due to its autoregressive nature. 
\cite{wang2022towards} (CSCv2) introduces a new non-autoregressive (NAR) contextual spelling correction model and incorporates contextual information into the decoder by attending to the contextual hidden representations from the bias encoder with an attention mechanism \cite{Attention-bahdanau2014}, as shown in Figure \ref{fig:csc_v3_model}. 
In CSCv2, the decoder directly takes hidden states from text encoder as input, attends to the bias phrase hidden representations, and outputs a position-wise classification (CLS) tag $cls$ and context index $cind$ for each input token. The CLS tag uses "BILO" representation where "B", "I", and "L" represent the beginning, inside and last position of a context phrase, "O" represents a general position outside of a context phrase; $cind$ is the expected index of the ground-truth context phrase in the bias list for each position. The final correction output can then be determined by $cls$ and $cind$. CSCv2 greatly improves the inference efficiency especially for low-end devices or resource limited systems but has similar biasing performance like its AR counterparts. 

However, both CSCv1 and CSCv2 may fail on the cases that hypotheses are totally irrelevant to the ground truth context phrase or on the cases that have more biasing phrases with similar pronunciation but dissimilar written format. On the other hand, although \cite{CSC, wang2022towards} use filter mechanisms for large context list to improve inference efficiency, its training hypotheses-reference pairs prepared with synthesized\cite{liu2021delightfultts} or human speech still have similarity gap in real scenario and it’s not easy for a CSC model to distinguish similar phrases with similar pronunciation or written format from limited hypothesis information only. 

Acoustics information has played an important role for ASR post processing besides text hypotheses in recent research \cite{ye2022have,hu2022transducer,le2022deliberation,peyser2020improving}. \cite{hu2020deliberation,hu2021transformer} combine both acoustics and first-pass text hypotheses for second-pass decoding, with an RNN-T or transformer model generating the first-pass hypotheses, then a deliberation model attending to both acoustics and first-pass hypotheses for a second-pass decoding. This shows improvement over text hypotheses only post processing model and inspires us to take acoustic information to improve CSC. The intuition is that if the text hypotheses can’t provide useful information for bias correction, acoustic information will help to complement the missing context.

In this paper, we proposed a new CSCv3 model, which combines both acoustic and text hypothesis for contextual spelling correction. Specifically, we introduce the designs as follows.
\begin{itemize}[leftmargin=*]
\item we propose to combine acoustics and first-pass text hypotheses for second-pass contextual spelling correction with biasing phrases as input. The proposed CSCv3 model has a similar structure as CSCv2 \cite{wang2022towards} except for the additional acoustic attention: a text encoder generates hidden vectors of the first-pass hypothesis from ASR model conditioned on acoustics hidden from ASR audio encoder output by an acoustic attention, a bias encoder generates contextual embeddings from biasing phrases, then a transformer decoder attends to both hypothesis and contexts for a second-pass decoding to generate phrase similarity and index for each encoder input position.
\item we use data augmentation for the context list construction during training with more similar pronunciation phrases or irrelevant phrases to improve the robustness of inference. 
\end{itemize}

Our experiments show that the proposed method achieves as much as 20.3\% relative name recall gain improvement compared to the baseline end-to-end ASR+Biasing system, and significantly outperforms the previous CSCv2 model.

\section{Methodology}
\label{sec:methodology}

\subsection{Model Architecture}

As shown in Figure \ref{fig:csc_v3_model}, the proposed CSCv3 model consists of 4 components: text encoder, bias encoder, acoustics adapter, and decoder. 
The text encoder takes ASR hypothesis as input and encodes text information into hidden states. 
The bias encoder converts the biasing phrases into phrase level embeddings, which adopts a multi-layer transformer encoder structure as in CSCv2.
The decoder uses the NAR (non-autoregressive) infrastructure which directly takes the encoder outputs as input and outputs a position-wise classification (CLS) tag $cls$ and context index tag $cind$ for each input token with the same definition described in Section \ref{sec:intro}

\subsubsection{External acoustic attention}

The acoustic encoder hidden of ASR model is adapted by the acoustic adapter in CSCv3 and then fed into CSC model with external attention. The acoustic adapter consists of two linear layers with ReLU activation function and dropout in between. 

We explore two different structures to leverage the acoustic information, as shown in Figure \ref{fig:csc_v3_model}. In CSCv3-EA, the acoustic information is incorporated into the text encoder by a cross attention which is added between the self-attention and feedforward module in each encoder layer, the decoder follows the same transformer decoder structure as in CSCv2, with self-attention, bias cross attention, and feedforward modules in each layer. For CSCv3-DA, the acoustic information is incorporated into the decoder by cross attention which is added between the self-attention and bias attention module in each decoder layer, while the text encoder keeps the same structure as in CSCv2. The comparison results between the two structures is shown in Section \ref{sec:results}.

Since the input audio feature sequence is typically long, we define an audio feature mask which masks audio features that are far from the position that corresponds to the current word piece token. For each token, we only attend to audio features corresponding to the surrounding $S_k$ words of this token. $S_k$ is randomly sampled from uniform distribution $[1, S_{kmax}]$, where $S_{kmax}$ is a pre-defined parameter.

\subsubsection{Semantic aware data augmentation}
\label{sec:anti}

The training pairs of CSCv3 are constructed randomly with the offline prepared data during training. To generate the biasing phrase list for each utterance, CSCv2 randomly samples $N_b$ biasing phrases from an existing large biasing phrase list besides the reference context phrase. $N_b$ is randomly sampled from uniform distribution $[1, N_{bmax}]$, where $N_{bmax}$ is the pre-defined max biasing phrase list size, which doesn’t consider irrelevant hypotheses or pronunciation similar phrases. 
As shown in the following table, CSCv3 improves the sampling strategy with the prepared reference-hypotheses pairs by two ways:

(1) Except for the raw ASR hypothesis of each utterance, we also randomly replace hypothesis of the context phrases with the prepared reference-hypotheses pairs, which improves the data varieties.

(2) To deal with anti-context cases where all biasing phrases are not present in the utterance, we randomly add two types of training data with probability $P_{anti}$: for the first type, the ground-truth context phrase is simply removed from the bias list, and the corresponding model output is also modified as non-context case; for the second type, we not only remove the ground-truth context phrase, but also add similar phrases into the biasing phrase list. These similar phrases come from the hypotheses in the prepared reference-hypotheses pairs.

\begin{table}[h]
  \vspace{-1.0em}
  \label{tab:training_pair_example}
  \centering
  \resizebox{0.85\linewidth}{!}{%
  \begin{tabular}{ll}
    \toprule
    Reference & \textit{Call \textcolor{blue}{John} at ten a.m.} \\
    ASR hypothesis & \textit{Call \textcolor{blue}{Joe} at ten a.m.} \\
    bias list & \{\textit{Sam}, \textit{John}, \textit{Dong}, ...\} \\
    Context ref-hyp pair & \textit{John} -- \{\textit{Jane}, \textit{Jon}, \textit{June}, \textit{Joe}\}  \\
    \midrule
    \multicolumn{2}{l}{(1) Replace hypothesis with ref-hyp pair:} \\
    Hyp $x$ & \textit{Call \textcolor{blue}{Jane} at ten a.m.} \\
    Ref $y$ & \textit{Call \textcolor{blue}{John} at ten a.m.} \\
    \midrule
    \multicolumn{2}{l}{(2.1) Remove ground-truth context phrase:} \\
    Hyp $x$ & \textit{Call \textcolor{blue}{Joe} at ten a.m.} \\
    bias list & \{\textit{Sam}, \sout{\textit{John}}, \textit{Dong}, ...\} \\
    Ref $y$ & \textit{Call \textcolor{blue}{Joe} at ten a.m.} \\
    \midrule
    \multicolumn{2}{l}{(2.2) Add similar phrases into bias list:} \\
    Hyp $x$ & \textit{Call \textcolor{blue}{Joe} at ten a.m.} \\
    bias list & \{\textit{Sam}, \sout{\textit{John}}, \textit{Dong}, \textit{Jane}, \textit{Jon}, ...\} \\
    Ref $y$ & \textit{Call \textcolor{blue}{Joe} at ten a.m.} \\
    \bottomrule
  \end{tabular}}
\vspace{-2.0em}
\end{table}

\subsubsection{Fast partial adaptation}

CSCv3 leverages both acoustics and text hypotheses information for better context biasing. Training from scratch is time consuming, for quick adaptation, we train the CSCv3 model based on a baseline CSCv2 model and only update new components in CSCv3, which includes audio adapter and acoustic attention layers. This strategy “inserts” acoustic information into the raw CSCv2 model and we will show its effectiveness in Section \ref{sec:results}. We also use a parameter $r$ to incorporate the acoustic information into the model. In each encoder layer of CSCv3-EA and decoder layer of CSCv3-DA, the data flow of acoustic attention layer can be expressed as:

\begin{equation}
x = x_0 + r \cdot {\rm dropout}({\rm AcousticAtt}({\rm norm}(x_0))),
\end{equation}
where $x_0$ and $x$ are the input and output of acoustic attention layer. $r$ is randomly sampled from a uniform distribution $[0.0, 1.0]$, which represents the incorporation ratio of acoustic information in the model.

\subsection{Training Optimization}

\subsubsection{Loss Objectives}

Like CSCv2, the loss function is the sum of CLS tag loss and context index loss:
\begin{equation}
L = H(\widehat{y_{cls}}, y_{cls}) + H(\widehat{y_{cind}}, y_{cind}).
\end{equation}

We also use teacher-student learning \cite{li2014learning, hinton2015distilling} to make the model smaller and more efficient.

\subsubsection{Data processing}

To generate the training data for CSCv3, we first decode the E2E ASR model for utterances with person names which are extracted from ASR model training set. The top-one hypothesis, audio encoder outputs, and forced alignment of the hypothesis and audio are needed for training. Then we locate and tag the positions of person names in each transcript, which is used to construct reference outputs during training.

Despite the raw ASR hypothesis, we also used a text to speech (TTS) system to generate synthetic audios for the person names. These synthetic audios are then fed into the ASR model to get hypotheses with more varieties. In this way we construct a set of reference-hypotheses pairs for the person names.

\subsection{Inference}

Like CSCv2, we use an edit distance-based relevance ranker (rRanker) to pre-select biasing phrases from the raw biasing phrase list and deal with the possible scalability issue:
\begin{equation}
W_r^j = -\frac{\min_{i}({\rm edit\_distance}(c_j, e_i))}{{\rm len}(c_j)},
\end{equation}
where $e_i$ is the segment cut off from input ASR hypothesis with the same length of the context phrase $c_j$ from the $i$-th word. The final relevance ranker weight is the minimum value of these edit distance normalized by the length of $c_j$.

The E2E ASR model decodes in a streaming way, we use the intermediate results and their corresponding decoding positions to estimate the rough alignment between audio and hypothesis. This alignment is then converted to the audio feature mask as model input.

\section{Experiment}
\label{sec:experiment}

\subsection{Data sets}

\noindent\textbf{Training set} We use a small set and a large set as the training data, which include 0.2 thousand (K) hours and 17K hours of Microsoft in-house en-US data respectively. We do a full decoding of the training data with the baseline E2E ASR model to get hypothesis and audio encoder hidden for CSCv3 training. 

\noindent\textbf{Test set} The test set consists of 12 Microsoft Teams meetings. Each meeting corresponds to a name list which consists of 600 person names, this list is expanded to a larger bias list with around 1500 phrases during inference. To evaluate the model performance on anti-context cases where the ground-truth name does not appear in the bias list, we also prepared 4 sets of bias lists with $25\%$, $50\%$, $75\%$, and $100\%$ name coverage for each meeting. All the data is anonymized with personally identifiable information removed.

\subsection{Model settings}

\noindent\textbf{ASR model} The baseline ASR model is a Conformer-Transducer (C-T) ~\cite{gulati2020conformer} model with the efficient low-latency implementation \cite{chen2020developing}, trained with 64K hours Microsoft anonymized data. The dimension of audio encoder output is 512 and we only use top-1 text hypothesis for CSCv3 input. 

\noindent\textbf{Teacher model} For the teacher model, each transformer layer contains a multihead-attention with 8 heads, and a 2048-dim feedforward layer. The text encoder, bias encoder and decoder all consist of 6 transformer blocks. The acoustic adapter consists of a 2048-dim feedforward layer followed by layer normalization. For text encoder of CSCv3-EA, an acoustic attention layer with 8 heads is inserted after the self-attention layer. While for the decoder of CSCv3-DA, an acoustic attention layer with 8 heads is sandwiched between the self-attention layer and biasing cross attention layer in each decoder block. 

\noindent\textbf{Student model} For the student model, the text encoder, bias encoder, and decoder all consist of 3 transformer blocks. The embedding dimension is set to be 192, all the multi-head attentions have 4 heads, and dimension of feedforward layers is 768. The feedforward layer in the audio feature adapter is composed of a 512-dimension and a 192-dimension linear layer.

\section{\uppercase{Results}}
\label{sec:results}

\begin{table*}[htb]
  \caption{Model performance with different bias list name coverage}
  \label{tab:model_perf}
  \centering
  \begin{tabular}{lllllllll}
    \toprule
    \multirow{2}{*}{Model}  & \multicolumn{2}{c}{25\% Coverage}  & \multicolumn{2}{c}{50\% Coverage} & \multicolumn{2}{c}{75\% Coverage} & \multicolumn{2}{c}{100\% Coverage}   \\
    \cmidrule(r){2-3}\cmidrule(r){4-5}\cmidrule(r){6-7}\cmidrule(r){8-9}
     & \multicolumn{1}{c}{Recall} & \multicolumn{1}{c}{WER} & \multicolumn{1}{c}{Recall} & \multicolumn{1}{c}{WER} & \multicolumn{1}{c}{Recall} & \multicolumn{1}{c}{WER} & \multicolumn{1}{c}{Recall} & \multicolumn{1}{c}{WER}   \\
    \midrule
    C-T      & $50.2$ & $12.5$ & $50.2$ & $12.5$ & $50.2$ & $12.5$ & $50.2$ & $12.5$   \\
    \midrule
    C-T+Biasing      & $58.0$ & $12.6$ & $59.0$ & $12.6$ & $61.4$ & $12.6$ & $64.1$ & $12.6$   \\
    \midrule
    \ +CSCv2      & $60.4$ & $12.7$ & $63.7$ & $12.6$ & $70.3$ & $12.6$ & $75.1$ & $12.6$   \\
    \midrule
    \ +CSCv3-EA-S0-nAnti-r1.0      & $58.0$ & $12.8$ & $60.4$ & $12.7$ & $70.1$ & $12.7$ & $74.3$ & $12.7$   \\
    \midrule
    \ +CSCv3-EA-S0-nAnti-r0.1      & $61.0$ & $12.6$ & $64.1$ & $12.6$ & $70.3$ & $12.6$ & $75.1$ & $12.6$   \\
    \midrule
    \ +CSCv3-EA-S0     & $61.8$ & $12.7$ & $64.7$ & $12.7$ & $71.3$ & $12.6$ & $75.3$ & $12.6$   \\
    \midrule
    \ +CSCv3-EA-full      & $61.8$ & $12.8$ & $64.9$ & $12.8$ & $72.3$ & $12.8$ & $76.9$ & $12.7$   \\
    \midrule
    \ \textbf{+CSCv3-EA}      & \textbf{62.7} & $12.7$ & \textbf{65.9} & $12.7$ & \textbf{72.7} & $12.7$ & \textbf{77.1} & $12.6$   \\
    \midrule
    \ +CSCv3-DA      & $62.7$ & $12.7$ & $64.1$ & $12.7$ & $71.3$ & $12.7$ & $75.9$ & $12.7$   \\
    \midrule
    \ +CSCv3-EA-student      & $62.7$ & $12.7$ & $65.3$ & $12.7$ & $72.3$ & $12.6$ & $77.7$ & $12.6$   \\
    \bottomrule
  \end{tabular}
\end{table*}

\noindent\textbf{Baseline} 
We use a C-T model as the baseline, and the C-T model with FST biasing \cite{OTF_rescore4} which uses the same biasing phrase list as a strong baseline (C-T+Biasing). In Table \ref{tab:model_perf}, we compare the name recall and WER of the models with different bias list name coverage. Where $s\%$ name coverage means there are $s\%$ of the ground truth names appear in the bias list while the rest are missing. We can see FST biasing has already achieved large name recall improvement compared to the C-T model.

\noindent\textbf{Data augmentation} It should be noted that CSCv3-EA-S0-nAnti-r0.1 and CSCv3-EA-S0-nAnti-r1.0 are the same model decoded with different parameters $r=0.1$ and $r=1.0$. It’s trained with the small training set and without anti-context cases mentioned in Section \ref{sec:anti}. It shows that a small incorporation ratio of acoustic information ($r=0.1$) leads to better performance. When the acoustic information is fully incorporated ($r=1.0$), the model performance becomes worse. We have investigated the decoding results and found that when $r$ is large, the model becomes more “biasing” and anti-context related errors are more likely to appear. This condition is not preferred because we hope the model be more stable on such cases. However, when anti-context data augmentation is added, as shown in CSCv3-EA-S0 which is trained with the same small set, this problem is gone and we find whether to add incorporation ratio $r$ during training or inference does not influence the decoding results much, and the name recall is also improved. Which indicates that data augmentation leads to more stable results and better performance.

\noindent\textbf{External acoustic attention} CSCv3-EA-S0 shows that the model achieves name recall gain even with a small training set. CSCv3-EA is trained with the large training set, which shows significant performance improvement when the training set becomes larger compared to CSCv3-EA-S0. The comparison of CSCv3-EA and CSCv3-DA indicates that incorporating the acoustic information into the text encoder achieves larger performance improvement. We also fully trained a CSCv3-EA model with all parameters updated with the large training set, which is called CSCv3-EA-full model. We observe that CSCv3-EA-full does not perform as well as CSCv3-EA which is partially trained. One of the reasons is that CSCv3-EA can still benefit from the baseline CSCv2 model which was trained with richer text-based data; another reason is that the current training set lacks general utterances without person names, which makes the model more biasing. It should be noted that CSCv2 shows limited improvement when the bias list name coverage is small (e.g., $25\%$ and $50\%$), one of the reasons is that it wrongly corrects some anti-context cases where the bias list does not contain the ground truth name but with some other phrases with similar pronunciation. With the external acoustic information, CSCv3 can deal with such issues and achieve stable improvement among different name coverage.

\noindent\textbf{Model size and latency} In Table \ref{tab:model_perf}, CSCv3-EA-student is the student model of CSCv3-EA, which shows similar performance compared to CSCv3-EA but with smaller model size. We also tested the latency of different models on the test set on a machine with
2.60GHz CPU using single thread regardless of baseline ASR model. The quantized onnx student model of CSCv2 is 5.4MB with 38.9ms latency per utterance, while CSCv3 is 6.2MB with 45.5ms per utterance, which indicates slight increase of model size and latency due to the external acoustic attention. Table \ref{tab:latency} lists the latency proportions of different components in CSCv3, which shows the most time-consuming component is bias encoder. To further reduce the latency in real usage, we propose to use a cache mechanism for bias phrase embeddings in continuous ASR recognition (e.g., meetings) by caching pre-calculated bias phrase embeddings. By using this mechanism, the latency of CSCv3 can be reduced to 20.9ms per utterance with cache size 1000.

\begin{table}[h]
  \vspace{-1.0em}
  \caption{Latency proportions of different components in CSCv3}
  \label{tab:latency}
  \centering
  \resizebox{0.9\linewidth}{!}{%
  \begin{tabular}{llll}
    \toprule
    Acoustics adapter & Text encoder & Bias encoder & Decoder \\
    \midrule
    10\% & 11.4\% & 68.1\% & 10.5\% \\    
    \bottomrule
  \end{tabular}}
  \vspace{-2.0em}
\end{table}

\section{Conclusion}

In this work, we propose an improved non-autoregressive (NAR) spelling correction model for contextual biasing in end-to-end transducer-based ASR systems with external acoustic attention and semantic aware data augmentation. The proposed model is proved to outperform the baseline ASR+Biasing system by as much as 20.3\% relative name recall gain and achieves stable improvement compared to the traditional CSC method over different bias list coverage ratio.

\bibliographystyle{IEEEbib}
\bibliography{strings,refs}

\end{document}